\documentclass[sigconf]{acmart}
\makeatletter
\def\@ACM@checkaffil{
    \if@ACM@instpresent\else
    \ClassWarningNoLine{\@classname}{No institution present for an affiliation}%
    \fi
    \if@ACM@citypresent\else
    \ClassWarningNoLine{\@classname}{No city present for an affiliation}%
    \fi
    \if@ACM@countrypresent\else
        \ClassWarningNoLine{\@classname}{No country present for an affiliation}%
    \fi
}
\makeatother

\usepackage[ruled]{algorithm2e}
\usepackage{multirow}
\usepackage{subfigure}
\usepackage{sistyle}
\SIthousandsep{,}
\usepackage{makecell}
\usepackage{multirow}
\usepackage{filecontents}
\usepackage{bbm}
\usepackage{pgfplots}
\usetikzlibrary{patterns}
\usepackage[inline,shortlabels]{enumitem}
\usepackage{natbib}
\usepackage{csquotes}
\usepackage{hyperref}
\usepackage{url}
\usepackage{CJKutf8}

\AtBeginDocument{%
  \providecommand\BibTeX{{%
    \normalfont B\kern-0.5em{\scshape i\kern-0.25em b}\kern-0.8em\TeX}}}

\AtBeginDocument{%
  \providecommand\BibTeX{{%
    \normalfont B\kern-0.5em{\scshape i\kern-0.25em b}\kern-0.8em\TeX}}}

\makeatletter
\g@addto@macro\normalsize{%
  \abovedisplayskip 2pt plus1pt 
  \belowdisplayskip 2pt plus1pt
  \abovedisplayshortskip  2pt plus1pt%
  \belowdisplayshortskip  2pt plus1pt
}
\makeatother

\setcopyright{acmcopyright}
\copyrightyear{2023}
\acmYear{2023}
\acmDOI{X.X}

\acmConference[KDD '23]{29TH ACM SIGKDD CONFERENCE ON KNOWLEDGE DISCOVERY AND DATA MINING}{August 06--10, 2023}{Long Beach, California}

\acmPrice{15.00}
\acmISBN{978-1-4503-XXXX-X/18/06}

\begin{document}
\begin{CJK*}{UTF8}{gbsn}

\title{Semantic-Enhanced Differentiable Search Index Inspired by Learning Strategies}

\author{Yubao Tang}
\affiliation{
 \institution{CAS Key Lab of Network Data Science and Technology, ICT, CAS}
 \institution{University of Chinese Academy of Sciences, Beijing, China}
}
\email{tangyubao21b@ict.ac.cn}

\author{Ruqing Zhang}
\authornote{Research conducted when the author was at the University of Amsterdam.}
\affiliation{
 \institution{CAS Key Lab of Network Data Science and Technology, ICT, CAS}
 \institution{University of Chinese Academy of Sciences, Beijing, China}
}
\email{zhangruqing@ict.ac.cn}

\author{Jiafeng Guo}
\authornote{Jiafeng Guo is the corresponding author.}
\author{Jiangui Chen}
\affiliation{
 \institution{CAS Key Lab of Network Data Science and Technology, ICT, CAS}
 \institution{University of Chinese Academy of Sciences, Beijing, China}
}
\email{{guojiafeng,chenjiangui18z}@ict.ac.cn}

\author{Zuowei Zhu}
\author{Shuaiqiang Wang}
\affiliation{
 \institution{Baidu Inc.}
 \city{Beijing}
 \country{China}
}
\email{{zhuzuowei,wangshuaiqiang}@baidu.com}

\author{Dawei Yin}
\affiliation{%
  \institution{Baidu Inc.}
  \city{Beijing}
  \country{China}}
\email{yindawei@acm.org}

\author{Xueqi Cheng}
\affiliation{
 \institution{CAS Key Lab of Network Data Science and Technology, ICT, CAS}
 \institution{University of Chinese Academy of Sciences, Beijing, China}
}
\email{cxq@ict.ac.cn}
\renewcommand{\shortauthors}{Yubao Tang et al.}

\begin{abstract}

Recently, a new paradigm called Differentiable Search Index (DSI) has been proposed for document retrieval, wherein a sequence-to-sequence model is learned to directly map queries to relevant document identifiers. 
The key idea behind DSI is to fully parameterize traditional ``index-retrieve'' pipelines within a single neural model, by encoding all documents in the corpus into the model parameters. 
In essence, DSI needs to resolve two major questions: (1) how to assign an identifier to each document, and (2) how to learn the associations between a document and its identifier. 
In this work, we propose a Semantic-Enhanced DSI model (SE-DSI) motivated by Learning Strategies in the area of Cognitive Psychology. 
Our approach advances original DSI in two ways: 
(1) For the document identifier, we take inspiration from Elaboration Strategies in human learning. Specifically, we assign each document an Elaborative Description based on the query generation technique, which is more meaningful than a string of integers in the original DSI;   
and (2) For the associations between a document and its identifier, we take inspiration from Rehearsal Strategies in human learning. 
Specifically, we select fine-grained semantic features from a document as Rehearsal Contents to improve document memorization.   
Both the offline and online experiments show improved retrieval performance over prevailing baselines. 

\end{abstract}



\maketitle


\section{Introduction}

Document retrieval is a fundamental task in many real-world applications, such as Web search and question answering systems \cite{application1,application2,application3}. 
It aims to identify a list of candidates from a large document repository given a user query. 
These candidates are then re-ranked to create a final list of results by computing a more precise ranking score for each document.  
The performance of the initial retrieval stage is crucial to the overall quality of the search systems. 
Traditional algorithms such as BM25 \cite{bm25} usually utilize exact term matching signals through the use of an inverted index. 
However, this method can run into issues with the vocabulary mismatch \cite{zhao2010term,furnas1987vocabulary} due to the independence assumption.

Major progress has recently turned to dense retrieval due to advances in deep learning especially representation learning techniques \cite{guo2022semantic}.  
These methods convert the semantic information in both queries and documents into dense vectors, and then use approximate nearest neighbor search algorithms \cite{aumuller2020ann} to perform efficient vector search \cite{khattab2020colbert}. 
Although dense retrieval has been shown to be effective in practical applications, the ``index-retrieval'' pipeline makes it difficult to jointly optimize all heterogeneous modules in an end-to-end way.  
Besides, an explicit large index is needed to conduct a search over the whole corpus, leading to significant memory consumption and computational overhead.

\begin{figure}[t]
 \centering
 \includegraphics[width=0.47\textwidth]{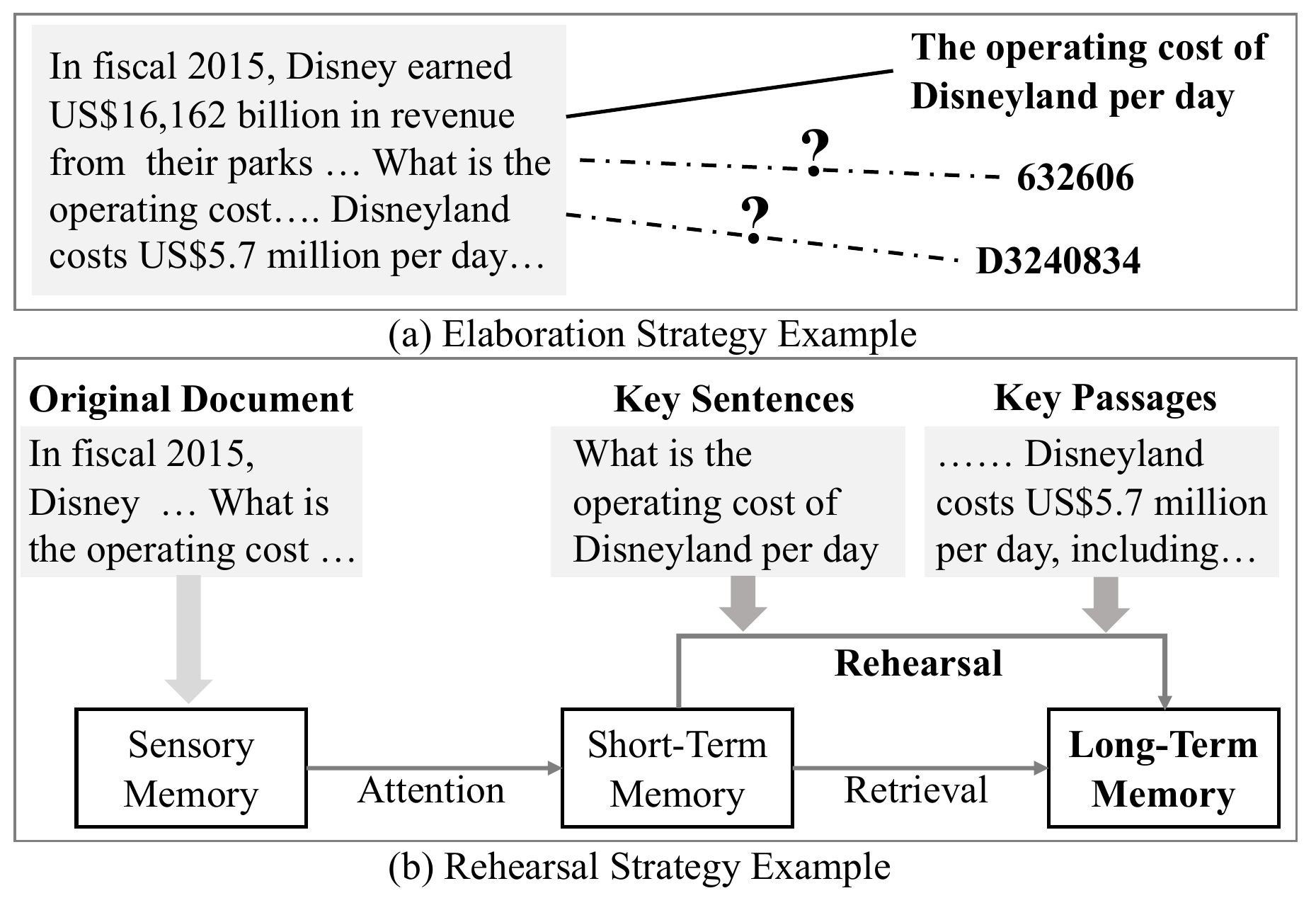}
 \caption{(a) Elaboration Strategies: Given a document, a semantically meaningful name, e.g., document title, could help people better encode and recall it than a weak-semantically meaningful name, e.g., a string of integers. (b) Rehearsal Strategies: By selectively underlining or highlighting the details in the document (e.g., key passages and sentences), people are more likely to ensure information goes from short-term memory to long-term memory than simply reading the document without underlining.}
 \label{fig:intro}
\end{figure}

Recently, \citet{DSI} proposed an alternative paradigm, called Differentiable Search Index (DSI). 
The key idea is to fully parameterize different components of index and retrieval with a single consolidated model, in which all information about the corpus is encoded in the model parameters. 
In essence, DSI adopts a generative scheme to directly predict the relevant document identifiers (docids) with a given query.  
DSI achieves this functionality by jointly optimizing two basic tasks: 
\begin{enumerate*}[label=(\roman*)]
    \item the indexing task, learning a mapping from the document content to its identifier (docid). The index is stored in model parameters, and indexing is simply another kind of model training.
    \item the retrieval task, mapping queries to relevant docids. 
\end{enumerate*} 
In this way, such a consolidated model can be optimized directly in an end-to-end manner towards a global objective. 
And DSI does not need to manage a complicated explicit index structure, largely reducing the memory and computational cost.

As envisioned in the recent proposal paper \cite{modelBased} and the original DSI \cite{DSI}, DSI needs to answer two major questions:   
(1) How to assign an identifier to each document, 
and then (2) How to learn the associations between a document and its identifier. 
As solved in \cite{DSI}, it used a single token (arbitrary unique integer) or a string of tokens which can be an arbitrary numeric string or a semantic numeric string via hierarchical clustering, as the docid. 
Besides, to bind a document to its docid, it utilized a straightforward seq2seq approach that takes the original documents as inputs and generates docids as outputs. 
Despite the superiority of the original DSI model over BM25 \cite{bm25} on the NQ 100K dataset \cite{naturalquestion},  
some follow-up studies \cite{bridging,NCI} and our work have shown that it still performs worse than state-of-the-art methods by a large margin. 
Such observation indicates that how to design a generative model for retrieval is still an open challenge for researchers.

 When we look at the process of corpus encoding in DSI, we find it works like that human uses interconnected ``neurons'' to learn to identify patterns in data and then directly make predictions about what should come next. 
Therefore, in this work, we resolve to design DSI models inspired by Learning Strategies \cite{weinstein1983learningstrategies} in Cognitive Psychology \cite{reisberg1997cognition,solso2005cognitive}. 
As defined in \cite{weinstein1983learningstrategies}, Learning Strategies are behaviors and thoughts in which a learner engages and which are intended to influence the learner's encoding process  \cite{morcom2003memoryencoding}. 
In a similar manner, we propose a novel Semantic-Enhanced DSI model, SE-DSI for short, to further optimize the solutions to the above two questions. 
Our approach advances original DSI in two ways: 

For the docids, we draw inspiration from Elaboration Strategies in human learning \cite{taevs2010semantic,hyde1969semanticelb2,anderson1979elaborativesemanticelb3,belmore1981imagerysemanticelb4,brown2006benesemanticelb5}. 
As shown in Figure \ref{fig:intro}(a), naming a document with natural language having semantic relationships with it, would contribute to better encoding and recall for humans than an integer-based string. 
Therefore, we construct Elaborative Description (ED) as the docid from each document to identify it with explicit semantic meaning. 
Specifically, we leverage the query generation technique to generate the pseudo query as ED from the corresponding document.

For associations between documents and their docids, we draw inspiration from Rehearsal Strategies in human learning   \cite{weinstein1983learningstrategies,levin1988elaboration,weinstein1982training,weinstein1977cognitive,weinstein2011self}. 
As shown in Figure \ref{fig:intro}(b), ones who underline important contents in a document are able to recall substantially more information and have higher long-term memory than ones who simply read the document without underlining.  
Therefore, we tailor-make two augmentation methods to generate Rehearsal Contents (RCs) at a different semantic granularity. 
The original document with coarse-grained semantic features and RCs with fine-grained semantic features can then be paired with the corresponding ED as training instances for better memorizing the documents.


Offline experiments on two representative document retrieval datasets, i.e., MS MARCO and NQ, show that the SE-DSI can perform significantly better than strong baseline solutions. 
We also simulate the zero-resource setting and show that SE-DSI works well even only with the document information.  
We also conduct an online evaluation on Baidu search\footnote{https://www.baidu.com} through A/B test. 
The results show that SE-DSI can achieve significant improvements over existing methods in Baidu on the official site retrieval task.

\section{Preliminaries}
\label{sec_prelimilary}

For a better description of our model, we first briefly describe the basic idea of the original DSI model \cite{DSI}, unifying two basic modes of operation, i.e., indexing and retrieval in an end-to-end way.

\textbf{Indexing:} To memorize information about each document, \citet{DSI} directly takes each original document $d_i$ as input and generates its docid $i$ as output in a straightforward Seq2Seq fashion. The model is trained with the standard T5 \cite{t5} training objective with the teacher forcing policy, i.e., 
\begin{equation*}
    \mathcal{L}_{index}(\theta) = \sum_{d_i \in \mathcal{D}} \log P(i|T5_{\theta}(d_i)), 
\end{equation*}
where  $\mathcal{D}$ is a given corpus and the docid $i$ could be represented by three ways, including, 
(1) atomic docid, wherein each document is assigned an arbitrary integer. 
Each docid is a single token in the T5 vocabulary and the decoder learns a probability distribution over the docid embeddings. 
However, it is difficult to apply such docid to large-scale corpus since the size of the model embedding layer cannot be too large. 
(2) string docid, wherein each document is assigned an arbitrary tokenizable numeric string. 
The decoder generates docids token-by-token in an autoregressive fashion.  
Such a way frees the limitation for the corpus size that comes with unstructured atomic docid. 
(3) semantic numeric docid, wherein a simple hierarchical clustering algorithm is employed over all the documents and each document is assigned an identifier with the number of their corresponding clusters. 
The experimental results in \cite{DSI} have also shown that the semantically structured docid performs better than the other two. 
However, all these integer-based docids have limited and implicit semantic meanings, which are not very consistent with human learning.

\textbf{Retrieval:} Given an input query $q$ in the query set $\mathcal{Q}$, a DSI model returns a docid by autoregressively generating the docid string $i$ with the fine-tuned T5 on indexing. The model is also trained with the standard T5 training objective,  
\begin{equation*}
    \mathcal{L}_{retrieval}(\theta) = \sum_{q_j \in \mathcal{Q}} \log P(i|T5_{\theta}(q_j)),
\end{equation*}
where $i$ is the generated docid for $q_j$.  A potentially-relevant ranked docids can be easily obtained with beam search \cite{medress1977speechbeam}.

\begin{figure*}[t]
\centering
\includegraphics[scale=0.63]{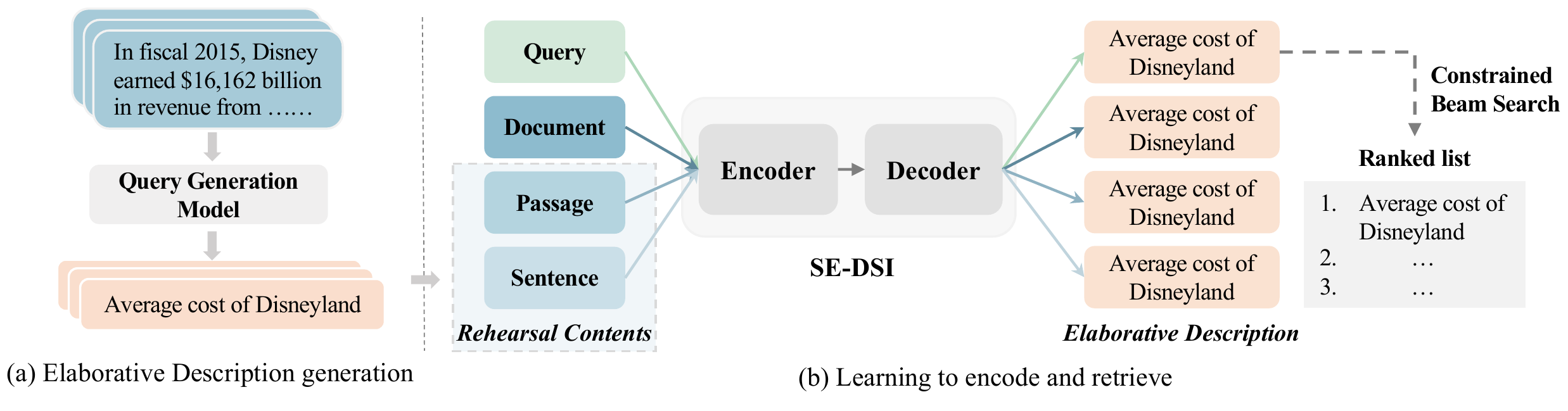}
\caption{An overview of our SE-DSI model. (a) We employ a query generation module to obtain ED from a document as its docid. 
(b) In the indexing phase, we propose to pair the original document and Rehearsal Contents (i.e., passage-level and sentence-level information) with the corresponding docid, respectively.  
In the retrieval phase, the docids are generated from the query, and a rank list of potentially-relevant documents is returned via beam search.}
\label{fig:diff-dsi}
\vspace*{-3mm}
\end{figure*}

\citet{DSI} proposes two main strategies for training DSI models. 
The first one is to first fine-tune T5 to perform indexing, followed by using the trained model for retrieval. 
The second one is to fine-tune T5 to perform both indexing and retrieval together in a multi-task setup.  
Through their experimental analysis, the second one performed significantly better.  
The multi-task learning is, 
\begin{equation*}
    \mathcal{L}_{DSI}(\theta) = \sum_{d_i \in \mathcal{D}} \log P(i|T5_{\theta}(d_i)) + \sum_{q_j \in \mathcal{Q}} \log P(i|T5_{\theta}(q_j)).
\label{dsi-loss}
\end{equation*}

Once such a DSI model is learned, it can be used to retrieve candidate documents for a test query $q_t$ in an end-to-end manner, 
\begin{equation*}
    i_p = DSI(q_t, i_0, i_1, \dots, i_{p-1}),
\label{dsi-loss}
\end{equation*}
where $i_p$ is the $p$-th token in the docid string and the generation stops when decoding a special EOS token. 
The generated string might not always be a valid docid if allowed to generate any token from the vocabulary at every decoding step. 
Hence, a constrained beam search strategy \cite{genre} is employed to force each generated docid string to be in a predefined candidate set.

    \vspace*{-2mm}
\section{Our Approach}
\label{sec_approach}

In this section, we introduce the SE-DSI model, a novel semantic enhanced DSI method designed for ad-hoc retrieval.   


    \vspace*{-2mm}
\subsection{Overview}

Formally, suppose $\mathcal{D}=\{d_1,d_2,...\}$ denotes a corpus, where $d_i$ is an individual document assigned a docid $i$. 
In DSI, docids are predicted using model parameters only.
This way, it shares a similar way to human recall or retrieval the information that was previously encoded and remembered in the brain \cite{fisher1980effectsmem,eysenck1979processingmem,pressley1982elaborationmem}. 
Therefore, we introduce a novel Semantic-Enhanced DSI model (SE-DSI) to advance original DSI, inspired by problem-solving strategies labeled by some psychologists, i.e., Learning Strategies \cite{huang2013learning,weinstein1983learningstrategies,reisberg1997cognition,weinstein2011self}.

Basically, the SE-DSI first constructs Elaborative Description (ED) from documents as docids to represent them with explicit semantics (Section \ref{method:ED}). 
Then, multiple coarse-fined contents from each document at different granularity are selected as Rehearsal Contents (RCs) (Section \ref{sec:RC}). 
In this way, we learn to build associations between original documents augmented with RCs and their corresponding EDs (Section \ref{method:train}). 
The overall architecture of SE-DSI is illustrated in Figure \ref{fig:diff-dsi}.

    \vspace*{-2mm}
\subsection{Elaborative description}
\label{method:ED}

Compared to designing an arbitrary integer or a string of integers as docids for documents, a more natural way for us humans is to describe the documents in natural language. 
In Elaboration Strategies, it is well known that for many memory tasks, learning with semantic elaboration, facilitates long-term memory and recall more than learning without semantic elaboration  \cite{belmore1981imagerysemanticelb4,levin1988elaboration,anderson1979elaborativesemanticelb3,hyde1969semanticelb2,brown2006benesemanticelb5}. 
Semantic elaboration can be defined as the process of stating a to-be-remembered stimulus, e.g., a story or picture, in natural language having semantic relationships with it, instead of non-nameable stimuli with weak semantics \cite{taevs2010semantic}. 
These motivate us to construct ED as the docids for documents.

It is intuitive that asking annotators to produce meaningful names for all documents in a large-scale corpus is time-consuming and requires increasingly sophisticated domain knowledge. 
To reduce the manual efforts of writing elaborative identifiers from scratch, we propose to generate ED by a query generation technique. 
Specifically, we leverage the off-the-shelf DocT5query model \cite{doct5query}, to generate pseudo queries as the docids, which are likely to be representative or related to the contents of documents. 
For each document $d_i$ in the given corpus $\mathcal{D}$, we directly feed it to the DocT5query model, to generate a set of representative queries with random sampling strategy. 
By conducting analysis on the two retrieval datasets used in this study, we find that concatenating more generated queries as the docid for generation,  leads to degraded retrieval performance. 
The possible reason is that the concatenated text is relatively longer than a query and a generative model is prone to hallucinate unintended content especially when the target sequence gets longer \cite{chen2020customizablelongtext,ji2022survey}.

In this work, we leverage the top 1 generated query as the ED for each document $d_i$, i.e., $ED_i$.  
Unfortunately, according to the experimental results, we find that about 5\% and 3\%  EDs of documents are not unique in MS MARCO and NQ respectively.  
It is reasonable that different documents may share the same ED if they share very similar essential information, which is similar to human learning: humans prefer to remember semantically similar documents with the same name. 
Following \cite{seal}, we ignore the ED repetition problem at the training phase.  
In the inference phase, since both datasets set the number of most ground-truth relevant documents as 1, we propose to solve the repetition problem in a simple way.  
Firstly, we leverage beam search to generate a ranked ED list.
Then, we obtain the corresponding documents of EDs to form the final ranked document list. 
If an ED corresponds to multiple documents, we return all of them in a random order, while keeping the relative order of documents corresponding to other EDs.

\vspace*{-3mm}
\subsection{Rehearsal contents}
\label{sec:RC}

To help ensure information goes from short-term memory to long-term memory, a very useful rehearsal strategy 
is to selectively underline or highlight multiple important parts when reading a new text~\cite{reisberg1997cognition}. 
This helps people reduce lengthy text into a comprehensible and manageable size that is central to understanding the piece and easy to memorize. 
Inspired by this learning strategy, we propose to select multiple important parts in a document as RCs to shorten the original document. And the original documents augmented with RCs are used to memorize the original document. Specifically, the RCs should fulfill the following conditions: 

 \textbf{Informative}: The RCs should contain the important information of the original document, enabling the model to learn to comprehend and encode the document into the parameters.  
 
\textbf{Fluency}: The RCs should be fluent and readable for the model to acquire the text encoding ability. 

\textbf{Diversity}: The RCs should contain different granularity of semantic units (e.g., the sentence- and passage-level), so as to achieve elaboration of the document for storage enhancement.

To achieve these goals, we propose to generate coarse-fined RCs at different granularity from each original document to rehearse it.  
Given a document, we select the important language units, i.e., passages and sentences, to condense it into RCs. 
Specifically, we tailor-make two data augmentation methods to generate RCs: 

\textbf{Leading-style}. We first introduce a simple but effective way to data augmentation method.
It is based on a simple fact: writers are likely to state major points at the beginning of the document and readers prefer to read the beginning part first.  
This leads to an intuitive idea: we can directly use the leading passages and sentences of each original document as its RCs. 
Specifically, for each document, we directly use the first $l$ passages and the first $k$ sentences as the passage- and sentence-level RCs, respectively. 

\textbf{Summarization-style}. We propose to incorporate the important information from the local context (e.g., sentence-level) and the broader context (e.g., paragraph-level). 
We leverage the document summarization technique to highlight multiple important parts that can reveal the essential topics of the document.  
We adopt a widely-used assumption, which denotes that a part is important in a document if it is highly related to many important parts \cite{zhang2018questioncikm}. 
We leverage a representative graph-based extractive summarization model TextRank~\cite{mihalcea2004textrank}, which uses co-occurrence information between words in the document to measure the importance of each part based on the PageRank \cite{langville2004deeperpagerank} algorithm. 
Specifically, for each document, we extract $n$ important passages and $u$ sentences as the passage- and sentence-level RCs, respectively.

Afterward, we can obtain a set of passage- and sentence-level RCs (denoted as $RC_i^p$ and $RC_i^s$, respectively) for each document $d_i \in \mathcal{D}$. 
The original document $d_i$ rehearsed by its RCs can then be paired with the $ED_i$ of $d_i$ as training instances to learn the mapping relationships between a document and its ED. 
Each RC shares the ED with the original document, contributing to enhancing the memorization of the document from multiple perspectives.

\vspace{-3mm}
\subsection{Training and inference}
\label{method:train}

In the training phase, given a corpus $\mathcal{D}$, a set of pairs $\{RC_i^p, ED_i\}$, $\{RC_i^s, ED_i\}$ and $\{d_i, ED_i\}$ for each document $d_i \in \mathcal{D}$, and the labeled query-ED pairs $\{q_j, ED_i\}$ for each $q_j$, 
we follow the multi-task learning strategy in the original DSI model, i.e.,  
\begin{equation*}
\begin{split}
        \mathcal{L}(\theta) = \sum_{d_i \in \mathcal{D}} log P(ED_i|SE_{\theta}(d_i)) +
                    \sum_{d_i \in \mathcal{D}} log P(ED_i|SE_{\theta}(RC_i^p)) + \\
                    \sum_{d_i \in \mathcal{D}} log P(ED_i|SE_{\theta}(RC_i^s)) +
                    \sum_{q_j \in \mathcal{Q}} log P(ED_i|SE_{\theta}(q_j)),
\end{split}
\label{dsi-loss}
\end{equation*}
where $SE$ denotes our SE-DSI model. 
To specify which task the model should perform (i.e., indexing and retrieval), we add a task-specific prefix ``Query'' to the input query $q_j$, and ``Document'' to the $RC_i^p$, $RC_i^s$ and $d_i$ before feeding it to the model.

In the inference phase, to ensure the decoded ED is valid, we employ a constrained Beam Search strategy \cite{medress1977speechbeam} to force each generated string to be in a pre-defined candidate set, i.e., the EDs of all the document in $\mathcal{D}$. 
Specifically, we define our constraint in terms of a prefix tree where nodes are annotated with tokens from the predefined candidate set.

\vspace*{-2mm}
\section{Offline Experimental Settings}
\label{sec_Experiment}

\subsection{Datasets}
\label{sub_Experiment_Dataset}
Following \cite{DSI,NCI,bridging}, we conduct offline experiments on two publicly available retrieval datasets, including, (1) \textbf{MS MARCO Document Ranking dataset (MS MARCO)} \cite{msmarco} is a large-scale benchmark dataset for web document retrieval.  Following~\cite{DSI}, to evaluate how models perform at different scales, we construct three sets from MS MARCO to form our testbed, namely MS MARCO 10K, MS MARCO 100K and MS MARCO Full. For MS MARCO 10K, we first randomly sample 14,763 and 1330 query-document pairs in the training set and dev set, respectively. Similarly, for MS MARCO 100K, we randomly sample query-document pairs from the training set and dev set, respectively. Besides, we refer to MS MARCO Full as the original dataset with about 3.21M documents. 
(2) \textbf{Natural Questions (NQ)} \cite{naturalquestion} contains 307K query-document pairs, where the queries are natural language questions and documents are gathered from the Wikipedia Pages. Following \cite{DSI},  we randomly sample 100,853 and 2800 query-document pairs in the training set and dev set to form \textbf{NQ 100K}.  
The dataset statistics are shown in Table \ref{tab:MS MARCO}. 
We use the original validation set of MS MARCO and NQ for evaluation following \cite{DSI,NCI,dai2020contextHDCT, ma2021b,ma2021pre}, since both MS MARCO and NQ leaderboard limit the frequency of submission.

\begin{table}[t]
    \caption{Statistics of datasets. \#Doc denotes the number of documents. \#Train denotes the number of the query-document pairs in training set. \#Dev denotes the number of queries in dev set. The dev set is used for evaluation.}
    \label{tab:MS MARCO}
    \centering
    \renewcommand{\arraystretch}{0.9}
    \setlength\tabcolsep{10pt}
    \begin{tabular}{lrrr}
         \toprule
        \textbf{Dataset} & \textbf{\#Doc}  & \textbf{\#Train} & \textbf{\#Dev} \\
        \midrule
       \textbf{MS MARCO 10K} & 13,569 & 14,763 & 1,330 \\
       \textbf{MS MARCO 100K} & 89,154 & 96,948 & 3,000\\
       \textbf{MS MARCO Full} & 3,213,835 & 367,013 & 5,193\\
       \textbf{NQ 100K} & 100,000 & 100,853& 2,800  \\
        \bottomrule
    \end{tabular}
\vspace{-2mm}
    
\end{table}

\vspace{-2mm}
\subsection{Evaluation metrics}
Following the original DSI model \cite{DSI} and some follow-up studies \cite{NCI,bridging}, we take Hit ratio (Hits@$N$) and Mean Reciprocal Rank (MRR@N) as the evaluation metrics.
Hits@N is the proportion of the right ranked document in the top $N$ ranking list, where $N$=\{1,10\}.
MRR calculates the reciprocal of the rank of the first $N$ retrieved relevant documents, where $N$=\{3,20\}.

\vspace{-2mm}
\subsection{Models}
\label{subsec_experiment_baseline}

\quad \textbf{Traditional document retrieval methods}. We consider two representative methods, including sparse retrieval and dense retrieval.
\begin{enumerate*}[label=(\roman*)]
    \item \textbf{BM25} \cite{bm25} is a term-based sparse retrieval method. We implement it with the Anserini open-source toolkit\cite{anserinitool}.
    \item \textbf{RepBERT} \cite{zhan2020RepBERT} is a BERT-based two-tower model trained with in-batch negative sampling. We implement it with the released code. We sample 1 negative sample for each positive sample. The batch size is 30 and learning rate is 1e-5. The max input length of the document and the query is 512 and 20, respectively. 
\end{enumerate*}

\textbf{DSI methods}. We also apply several existing DSI methods.
For docids described in Section \ref{sec_prelimilary}, we consider the unique arbitrary string and semantic numeric string.
Since the effect of the single token is worse than these two ones, reported in \cite{DSI}, we ignore this type. 
For the indexing strategy, we choose two effective methods, including learning (document, docid) pairs and (pseudo query, docid) pairs, reported in \cite{DSI, NCI, bridging}.
For the implementation of DSI methods, we use the same settings as our SE-DSI model. 
\begin{enumerate*}[label=(\roman*)]
    \item \textbf{DSI-ARB} takes the original documents as input and outputs the corresponding unique ARBitrary string docids in \cite{DSI}. 
    \item \textbf{DSI-SEM} takes the original documents as input and outputs the corresponding SEMantic numeric string docids in \cite{DSI}. 
    \item  \textbf{DSI-QG} takes a set of pseudo Queries Generated by the original documents with a query generation model as input, and outputs semantic numeric docids. 
    It can be viewed as the adaption of \cite{NCI, bridging}. 
\end{enumerate*}

\textbf{Model variants}. We refer to our SE-DSI model with leading- and summarization-style augmentation methods as \textbf{SE-DSI$_{Lead}$} and \textbf{SE-DSI$_{Sum}$}, respectively. 
We also implement two variants of SE-DSI, namely \textbf{SE-DSI$_{Doc}$} and \textbf{SE-DSI$_{Random}$}. 
\textbf{SE-DSI$_{Doc}$} takes as input the original document and outputs its ED.
\textbf{SE-DSI$_{Random}$} achieve RCs by randomly sampling several passages and sentences from the document, where the number of passages and sentences follows the leading-style augmentation method.

\vspace*{-2mm}
\subsection{Implementation Details}\label{sec:implemnetation}

\quad \textbf{Elaborative Description}. For MS MARCO, we use the released pseudo queries generated by docT5query\cite{doct5query} as ED. 
For NQ, following ~\cite{NCI}, we directly leverage the docT5query model to generate 10 queries for each document. 
The maximum length of a pseudo query is fewer than 20 for both MS MARCO and NQ. 

\textbf{Rehearsal Contents}. We first split each document by spacy's sentencizer \cite{sentencizer}.
Following~\cite{doct5query}, we regard 5 successive sentences as one passage and skip two sentences to obtain the next passage. 
After iterating in this way, we can obtain a sequence of passages. 
According to our statistics, the percentage of documents with fewer than 3 passages is 3\% in MS MARCO, and 4\% in NQ. 
For the leading-style augmentation method in RCs, we set the number of the leading passages $l$ and the leading sentences $k$ to 3 and 6, respectively. 
Note for the document with fewer than 3 passages, we set $l$ as 1, while for the document with fewer than 6 sentences, we use all the sentences. 
For summarization-style augmentation method, we set the number of important passages $n$ and important sentences $u$ as 1 and 6, respectively. 
Specifically, we leverage the summa  API \cite{summa} to implement the TextRank model.

\textbf{Training and Inference}. Since the original code is not publicly available by the authors \cite{DSI}, we implement and train our model and existing DSI models by ourselves. 
We employ the Transformer-based encoder-decoder architecture as our model, where the hidden size is 768, the feed-forward layer size is 12, the number of self-attention heads is 12, and the number of Transformer layers is 12.
We initialize the parameters of our model with T5-base(0.2B)\cite{t5base}. 
Note existing DSI methods are also based on T5-base. 
We use Adam optimizer with a linear warm-up over the first 10\% steps.
The learning rate is set to 5e-5, the label smoothing is 0.1, the weight decay is 0.01, the sequence length is 512, the max training steps is 50K and the batch size is 30.
We train our model on four NVIDIA Tesla A100 40GB GPUs.
At inference time, we adopt constrained beam search to decode the ED with 20 beams. 

\begin{table*}[t]
\small
    \caption{Experimental results on the MS MARCO dataset. $\ast$, $\dagger$ and $\ddagger$ indicate statistically significant improvements over the best performing generative retrieval baseline DSI-QG, BM25, and RepBERT, respectively ($p \leq 0.05$).}
    \label{tab:msdoc-res}
    \centering
    \setlength{\tabcolsep}{3pt}
    \renewcommand{\arraystretch}{0.95}
    \begin{tabular}{l|cccccccccccccccc}
        \toprule
        \multirow{2}{*}{\textbf{Methods}} &  \multicolumn{4}{c}{\textbf{MS MARCO 10K}} &  \multicolumn{4}{c}{\textbf{MS MARCO 100K}} &  \multicolumn{4}{c}{\textbf{MS MARCO Full}}\\
        &  \textbf{MRR@3} & \textbf{MRR@20} & \textbf{Hits@1} & \textbf{Hits@10}  & \textbf{MRR@3} & \textbf{MRR@20} & \textbf{Hits@1} & \textbf{Hits@10} & \textbf{MRR@3} & \textbf{MRR@20} & \textbf{Hits@1} & \textbf{Hits@10} \\
        \midrule

        BM25 & 0.4049 & 0.4230 & 0.3760 & 0.5866& 0.3815 & 0.3700 & 0.4846 &  0.5363&  0.1784&  0.2168&0.1186&0.4358\\
        RepBERT & 0.4304 & \textbf{0.4776} & 0.4070  & 0.5874  & 0.4191 & 0.4459 & 0.4917 & 0.6195 & 0.2671 & 0.3078  & 0.1930 &	\textbf{0.5584} \\
        \midrule
        DSI-ARB &  0.1069 &	0.1274 &	0.1087 	&0.1377 &	0.1153 &0.1176 &0.1187&	0.1180 	&0.1053 &0.1079&0.1022& 0.1138 \\
        DSI-SEM &  0.2096 &	0.2152 &0.2045& 0.2392 &0.2103 &0.2196 &0.2054 &0.2544& 0.1331&	0.1479&	0.1092&	0.1678 \\
        DSI-QG &  0.4237 &	0.4497 &0.3831 &	0.5913 &	0.3997 &0.4233 &	0.3515 &0.5703 & 0.2277 &	0.2312 &	0.1980 	&0.2805  \\
                \midrule
  
        SE-DSI$_{Doc}$ &  0.2559 &0.2631 &	0.2360 &	0.3205 	&0.4686\rlap{$^{\ast\dagger\ddagger}$}	&0.4757\rlap{$^{\ast\dagger\ddagger}$}	&0.4360\rlap{$^{\ast}$} 	&0.5427 &0.2429\rlap{$^{\ast\dagger}$} 	&0.2516\rlap{$^{\ast\dagger}$} 	&0.2036\rlap{$^{\dagger}$} 	&0.3347\rlap{$^{\ast}$} \\
        
        SE-DSI$_{Random}$ &  0.4217\rlap{$^{\dagger}$}	&0.4425\rlap{$^{\dagger}$} &	0.3725 &	0.5837 &	0.4693\rlap{$^{\ast\dagger\ddagger}$} &	0.4819\rlap{$^{\ast\dagger\ddagger}$} &0.4320\rlap{$^{\ast}$}& 0.5774\rlap{$^{\dagger}$} & 0.2577\rlap{$^{\ast\dagger}$} &	0.2616\rlap{$^{\ast\dagger}$} &	0.2161\rlap{$^{\ast\dagger\ddagger}$} &	0.3561\rlap{$^{\ast}$}  \\
     \midrule
        \textbf{SE-DSI$_{Lead}$} &  0.4343\rlap{$^{\ast\dagger}$} &0.4582\rlap{$^{\dagger}$} &0.3876\rlap{$^{\dagger}$} &	\textbf{0.6063}\rlap{$^{\ast\dagger\ddagger}$} &	0.5171\rlap{$^{\ast\dagger\ddagger}$} 	&0.5314\rlap{$^{\ast\dagger\ddagger}$} &	0.4680\rlap{$^{\ast}$} &0.6478\rlap{$^{\ast\dagger\ddagger}$} &0.2779\rlap{$^{\ast\dagger\ddagger}$} &	0.2845\rlap{$^{\ast\dagger}$} 	&0.2381\rlap{$^{\ast\dagger\ddagger}$} &	0.3597\rlap{$^{\ast}$} \\
        
        \textbf{SE-DSI$_{Sum}$} & \textbf{0.4377}\rlap{$^{\ast\dagger}$} &	0.4567\rlap{$^{\dagger}$} & \textbf{0.4074}\rlap{$^{\ast\dagger}$} & 0.5830 & \textbf{0.5900}\rlap{$^{\ast\dagger\ddagger}$} & \textbf{0.6092}\rlap{$^{\ast\dagger\ddagger}$} & \textbf{0.5347}\rlap{$^{\ast\dagger\ddagger}$} & \textbf{0.7528}\rlap{$^{\ast\dagger\ddagger}$} & \textbf{0.3022}\rlap{$^{\ast\dagger\ddagger}$}&	\textbf{0.3463}\rlap{$^{\ast\dagger\ddagger}$}&	\textbf{0.2609}\rlap{$^{\ast\dagger\ddagger}$}&	0.4002\rlap{$^{\ast}$} \\

        \bottomrule
    \end{tabular}
    \vspace*{-2mm}
\end{table*}

\vspace{-1mm}
\section{Offline Experimental Results}

\subsection{Main results}

The comparison between our SE-DSI and baselines on MS MARCO and NQ 100K datasets is shown in Table \ref{tab:msdoc-res} and Table~\ref{tab:nq-res}. 

\textbf{Performance of sparse retrieval and dense retrieval methods}: 
(1) BM25 is a strong baseline that performs pretty well on most datasets. 
By automatically learning text representations and semantic relationships between queries and documents, RepBERT can achieve better results than BM25. 
(2) The performance gap gets larger as the size of the dataset increases. 
The reason might be that the dense retrieval methods trained with more data can improve the performance. However, the performance of BM25 does not change regularly with the size of the dataset.

\begin{table}[t]
    \caption{Experimental results on the NQ 100K dataset. $\ast$, $\dagger$ and $\ddagger$ indicate statistically significant improvements over the best performing generative retrieval baseline DSI-QG, BM25, and RepBERT, respectively ($p \leq 0.05$).} 
    \label{tab:nq-res}
    \centering
    \renewcommand{\arraystretch}{1.1}
    \begin{tabular}{l|cccc}
    
        \toprule
        \textbf{Methods} &  \textbf{MRR@3} & \textbf{MRR@20} & \textbf{Hits@1} & \textbf{Hits@10}  \\
        \midrule

        BM25 & 0.1846  &	0.1873  &	0.1742  &	0.2111 \\
        RepBERT & 0.3254  &	0.3339  &	0.2993  &	\textbf{0.5042}  \\
        \midrule
        DSI-ARB &  0.2224 	 &0.2684  &	0.2617 	 &0.3246 \\
        DSI-SEM &   0.2516  &	0.2801 	 &0.2699  &	0.3427 \\
        DSI-QG &    	0.3131 & 	0.3220 & 	0.2903  &	0.3869 \\
        \midrule
  
        SE-DSI$_{Doc}$ &  0.2916\rlap{$^{\dagger}$}  &	0.3001\rlap{$^{\dagger}$} 	 &0.2700\rlap{$^{\dagger}$}  &	0.3627\rlap{$^{\dagger}$} \\
        SE-DSI$_{Random}$ &  0.3046\rlap{$^{\dagger}$}  &	0.3160\rlap{$^{\dagger}$}  &	0.2866\rlap{$^{\dagger}$}  &	0.3709\rlap{$^{\dagger}$}  \\
     \midrule
        \textbf{SE-DSI$_{Lead}$} & 0.3224\rlap{$^{\dagger}$} &	0.3329\rlap{$^\dagger$}  &	0.3078\rlap{$^\ast\dagger$} 	 &0.4087\rlap{$^\ast\dagger$}  \\
        \textbf{SE-DSI$_{Sum}$} & \textbf{0.3511}\rlap{$^{\ast\dagger\ddagger}$}  &	\textbf{0.3644}\rlap{$^{\ast\dagger\ddagger}$}  &	\textbf{0.3383}\rlap{$^{\ast\dagger\ddagger}$}  &	0.4555\rlap{$^{\ast\dagger}$}  \\

        \bottomrule
    \end{tabular}
\end{table}

\textbf{Performance of DSI baselines}:
(1) DSI-ARB and DSI-SEM perform better than BM25 on NQ 100K, which is consistent with the results in the original model \cite{DSI}. 
However, in accordance with some follow-up studies \cite{bridging}, DSI-ARB and DSI-SEM perform worse than sparse retrieval and dense retrieval baselines by a large margin on MS MARCO. 
The reason might be that it is hard for the model to learn associations between documents and integer-based string identifiers with limited semantic information. 
This again indicates that the performance of the DSI still has a large room for improvement. 
(2) The performance improvements of DSI-SEM over DSI-ARB, indicating imbuing the target space with semantic structure can facilitate greater ease of optimization \cite{DSI}. 
(3) The performance improvements of DSI-QG over DSI-SEM, show that bridging the gap of input data between indexing and retrieval helps the model better learn the association between query and docid. 
However, documents usually contain rich semantics and it may not be optimal to only encode pseudo queries and ignore documents.

\textbf{Performance of our SE-DSI}: (1) SE-DSI$_{Random}$ performs better than SE-DSI$_{Doc}$ significantly on all the datasets.
Besides the original document, SE-DSI$_{Random}$ also introduces randomly sampled passages and sentences, which does help enhance the document memorization. 
This result demonstrates that the corpus encoding process in DSI is similar to the rehearsal strategy to a certain extent. 
(2) SE-DSI$_{Sum}$ can outperform the baseline methods in terms of almost all the metrics, showing that employing ECs and EDs simulating the human learning process, can better contribute to indexing and retrieval.
(3) Our method performs worse than RepBERT on MS MARCO Full and NQ 100K in terms of Hits@10. 
The reason might be that RepBERT leverages the pair-wise loss considering the relationship between a positive and a negative document, while SE-DSI directly learns the query-ED relationship (but this helps it performs the best in terms of Hits@1).  
(4) Among the two of our models, SE-DEI$_{Sum}$ outperforms $_{Lead}$, indicating that important sentences and passages contain more useful information for document memorization than leading contents.

\textbf{Memory and inference efficiency}: SE-DSI$_{Sum}$ has a significant reduction of memory footprint and inference time of document retrieval compared to dense retrieval models.
\begin{enumerate*}[label=(\roman*)]
\item The major memory computation of SE-DSI$_{Sum}$ is a prefix tree of the document identifiers and the number of model parameters, as opposed to a large document index and a dense vector for each document in dense retrieval. For example, the memory footprint of our model is reduced by about 31 times compared to RepBERT.
\item The heavy retrieval process is replaced with a light generative process over the prefix tree, instead of the time-consuming step of searching over a large-scale corpus. For example, the inference speed of SE-DSI$_{Sum}$ is significantly improved by about 2.5 times compared to RepBERT.
\end{enumerate*}
Other variants of SE-DSI have the same phenomenon.

\begin{table}[t]
    \caption{An example from the MS MACRO 100K dev set. Given a query (QID:320792), which is relevant to D324083, SE-DSI$_{Doc}$ and DSI-SEM return the top-5 beams. Correct results are marked bold.}
    \label{tab:ab-case-id}
    \setlength{\tabcolsep}{4pt}
    \renewcommand{\arraystretch}{1}
    \begin{tabular}{lcccc}
        \toprule
        \multicolumn{3}{l}{\multirow{3}{*}{\begin{minipage}{0.95\linewidth}\textbf{Doc(D3240834)}: In 2015, Disney earned US\$16,162 billion... the operating cost of a single theme park is likely to be... it spends a lot on a daily basis, that could easily be 15-20\% ...\end{minipage}}} \\ \\ \\ \hline
        
      \multicolumn{3}{l}{\multirow{1}{*}{\begin{minipage}{0.95\linewidth}\textbf{Semantic Numeric Docid}: 632606\end{minipage}}} \\ \hline
    \multicolumn{3}{l}{\multirow{1}{*}{\begin{minipage}{0.90\linewidth}\textbf{Elaborative Description}: Average cost of Disneyland\end{minipage}}} \\ \hline
      \multicolumn{3}{l}{\multirow{1}{*}{\begin{minipage}{0.95\linewidth}\textbf{Query}: How much is a cost to run Disneyland? \end{minipage}}} \\  \hline
        \textbf{\#} & \textbf{DSI-SEM} & \textbf{SE-DSI$_{Doc}$} \\
        \hline
        1 &  632600 &   Cost of Disneyland tickets\\
        2 &  632605 &   Admission rate for Disneyland\\
        3 &  632604 & Disney ticket price\\
        4 &  632602 &  \textbf{Average cost of Disneyland}\\
        5 &  632603 & Cost of locker at Disneyland\\
       \bottomrule
    \end{tabular}
    \vspace{-2mm}
\end{table}

\begin{table*}[t]
    \caption{Experimental results of zero-shot retrieval settings on MS MARCO 100K and NQ 100K. $\ast$ indicates statistically significant improvements over the best performing baseline DSI-QG ($p \leq 0.05$).}
    \label{tab:zero}
    \centering
    \setlength{\tabcolsep}{10pt}
    \renewcommand{\arraystretch}{0.8}
    \begin{tabular}{l|cccc|cccccccccccc}
    
        \toprule
       \multirow{2}{*}{\textbf{Methods}} &   \multicolumn{4}{c}{\textbf{MS MARCO 100K}} &  \multicolumn{4}{c}{\textbf{NQ 100K}}\\
        &  \textbf{MRR@3} & \textbf{MRR@20} & \textbf{Hits@1} & \textbf{Hits@10}  & \textbf{MRR@3} & \textbf{MRR@20} & \textbf{Hits@1} & \textbf{Hits@10} \\
       
        \midrule
        DSI-ARB&	0.1044& 	0.1135& 	0.1016& 	0.1154 &	0.1345 &	0.1380 &	0.1282 &	0.1613 \\
        DSI-SEM&	0.1396& 	0.1545& 	0.1410& 	0.1621 &	0.1458& 	0.1507 	&0.1365 &	0.1833 \\
        DSI-QG&	0.2668 &	0.2725 &	0.2468& 	0.3193 & 0.2391 &	0.2446 &	0.2019 &	0.2836 \\
    	\midrule
        SE-DSI$_{Doc}$&	0.2631 & 	0.2700 & 	0.2420  	&0.3258  &	0.1923 &	0.2043 &	0.2116 &	0.2660  \\
        SE-DSI$_{Random}$&	0.2826\rlap{$^\ast$} & 	0.2903\rlap{$^\ast$} & 	0.2599\rlap{$^\ast$} 	&0.3505\rlap{$^\ast$} &	0.2285 &	0.2320& 	0.2217\rlap{$^\ast$} &	0.2813   \\
        \midrule
        \textbf{SE-DSI$_{Lead}$}&	0.3022\rlap{$^\ast$} &	0.3118\rlap{$^\ast$} &	0.2759\rlap{$^\ast$} &	0.3804\rlap{$^\ast$}   &	0.2430 &	0.2517 &	0.2285\rlap{$^\ast$} &	0.3077\rlap{$^\ast$} \\
        \textbf{SE-DSI$_{Sum}$}&	\textbf{0.4472}\rlap{$^\ast$}& 	\textbf{0.4326}\rlap{$^\ast$}& 	\textbf{0.4896}\rlap{$^\ast$} 	&\textbf{0.5564}\rlap{$^\ast$} &	\textbf{0.2900}\rlap{$^\ast$} &	\textbf{0.2947}\rlap{$^\ast$} &	\textbf{0.2672}\rlap{$^\ast$} &	\textbf{0.3405}\rlap{$^\ast$} \\
        \bottomrule
    \end{tabular}
    \vspace{-3mm}
\end{table*}

\begin{table}[t]
    \caption{Impact of different RCs on MS MARCO 100K. $\ast$ indicates statistically significant improvements over the best performing variant w/ Doc+Psg ($p \leq 0.05$).}
    \label{tab:ab-stepping-stone}
    \centering
    \setlength{\tabcolsep}{5pt}
    \renewcommand{\arraystretch}{0.9}
    \begin{tabular}{l|c|c|c|c}
        \toprule
        \textbf{Methods} &  \textbf{MRR@3} & \textbf{MRR@20} & \textbf{Hits@1} & \textbf{Hits@10} \\
        \midrule
         w/ Document	&		0.4686 &	0.4757 	&0.4360 &	0.5427 \\
        w/ Sentence	&		0.4326 &	0.4520 	&0.3813 &	0.5930 \\
        w/ Passage		&	0.3061 	&0.3143 &	0.2799 &	0.3781 \\
        \midrule
        w/ Doc+Sent	&	0.4702 &	0.4611 &	0.4844 &	0.6140 \\
        w/ Doc+Psg	&		0.4895	&0.500	&0.4503	&0.5884\\
        \midrule
      
         SE-DSI$_{Sum}$ &\textbf{0.5900}\rlap{$^\ast$} & \textbf{0.6092}\rlap{$^\ast$} &\textbf{0.5347}\rlap{$^\ast$} &\textbf{0.7528}\rlap{$^\ast$} \\

        \bottomrule
    \end{tabular}
    \vspace{-2mm}

\end{table}

\vspace{-2mm}
\subsection{Analysis on elaborative description}

In this section, we compare the proposed EDs to existing integer-based docids. 
As shown in Table \ref{tab:msdoc-res} and Table~\ref{tab:nq-res}, we can find that SE-DSI$_{Doc}$ performs better than DSI-ARB and DSI-SEM on both MS MARCO and NQ 100K.
These results indicate the effectiveness of representing a document with our proposed ED as the docid, which is a natural language text containing enhanced semantic meanings.

\textbf{Case}. We conduct case studies to see how EDs as docids affect performance. 
Specifically, we take one example from the MS MACRO 100K dev set, and show the top-5 retrieval results by SE-DSI$_{Doc}$ and DSI-SEM, which uses EDs and semantic numeric docids, respectively. 
As shown in Table \ref{tab:ab-case-id}, we can see that:
Given the same query, SE-DSI$_{Doc}$ ranks the ground-truth documents at the 4$-th$, while DSI-SEM can not rank it in top 5 (actually 10$-th$). 
Since the semantic numeric docid, i.e., ``63260'', is hard to reflect the semantics of the document, while ED as the docid, i.e., ``Average cost of Disneyland'' is easier to be representative of the document.

\vspace{-3mm}
\subsection{Analysis on rehearsal contents}

Here, we analyze whether RCs can help document memorization compared to the existing method which only takes the original document as the input on MS MARCO 100K. 
Specifically, for each document, firstly, we only feed the SE-DSI with the documents, the sentences, and the passages, respectively. 
Then, we feed the SE-DSI with the mixture of the documents and sentences, and that of the documents and passages, respectively. 
Here, we obtain the sentences and passages via the summarization way.

As shown in Table \ref{tab:ab-stepping-stone}, we can see that:
(1) Rehearsing the original documents with two granularity, i.e., w/ Doc+Sent and w/Doc+Psg, outperforms that with only one granularity, i.e., w/Doc, w/Psg and w/Sent. 
This indicates that it is insufficient to only encode the document content with single granularity. 
(2) The better results of w/Sent over w/Psg denotes that reducing the gap of input format between indexing and retrieval contributes to the final performance. 
However, both of them can not outperform w/doc, due to the loss of rich semantics in documents. 
(3) SE-DEI$_{Sum}$ achieves the best results, again indicating that our method learning with the underlined important contents of the documents 
can comprehensively encode the documents, and further contribute to the retrieval.

\textbf{Case}. We also conduct some case studies to better understand how RCs affect the performance. 
We take the document (D3240834) in Table \ref{tab:ab-case-id} as an example, and show the predicted EDs from SE-DSI$_{Sum}$ and  SE-DSI$_{Doc}$, which encode the documents in different ways, i.e., RCs and original documents, respectively. 
As shown in Table \ref{tab:ab-case-ec}, we can observe that:
Given the query, SE-DSI$_{Sum}$ and SE-DSI$_{Doc}$ rank the ground truth at the 1$-th$ and 4$-th$, respectively. 
This result shows that augmenting key information does help document memorization and distinguish similar documents.

\begin{table}[t]
    \caption{For the same  document (D3240834) in Table 4, EC-passage and EC-sentence are key passages and sentences of the document. Given the query, SE-DSI$_{Sum}$ and SE-DSI$_{Doc}$ return the top-5 beam. Correct results are marked bold.}
    \label{tab:ab-case-ec}
    \setlength{\tabcolsep}{2pt}
    \renewcommand{\arraystretch}{1}
    \begin{tabular}{lcccc}
        \toprule
      \multicolumn{3}{l}{\multirow{3}{*}{\begin{minipage}{0.95\linewidth}\textbf{EC-passage}: Disney's Theme Parks had an operating cost of 571 million dollars divided by their 11 parks and being open 365 days a year, on average their operating cost per day…
\end{minipage}}} \\ \\   \\\hline
    \multicolumn{3}{l}{\multirow{2}{*}{\begin{minipage}{0.95\linewidth}\textbf{EC-sentence}: How much does it cost Disney to run Disneyland per day including California Adventure Disney? 
\end{minipage}}} \\  \\ \hline

      \multicolumn{3}{l}{\multirow{1}{*}{\begin{minipage}{0.95\linewidth}\textbf{Query}: How much is a cost to run Disneyland? \end{minipage}}} \\ \hline
        \textbf{\#} & \textbf{SE-DSI$_{Sum}$} & \textbf{SE-DSI$_{Doc}$} \\
        \hline
        
        1 & \textbf{Average cost of Disneyland} & Cost of Disneyland tickets \\
        2 & Cost of Disneyland tickets & Admission rate for Disneyland \\
        3 & Cost of locker at Disneyland & Disney ticket price \\
        4 & Disney ticket price & \textbf{Average cost of Disneyland} \\
        5 & Admission rate for Disneyland & Cost of locker at Disneyland \\

       \bottomrule
    \end{tabular}
    \vspace{-1mm}

\end{table}

\vspace{-3mm}
\subsection{Zero-shot setting}

We further conduct zero-shot retrieval on MS MARCO 100K and NQ 100K. 
For a fair comparison, we only compare our model with existing DSI methods. 
Specifically, zero-shot retrieval is performed by only performing indexing without the retrieval task \cite{DSI}, i.e. the ground-truth query-document pairs are not provided in the training phase. 
As shown in Table \ref{tab:zero}, we can observe that:
(1) DSI-QG slightly outperforms SE-DSI$_{Random}$ on NQ 100K. 
That is probably because DSI-QG takes as input the pseudo-queries in indexing, which is similar to the input data in retrieval. 
(2) SE-DSI$_{Sum}$ can outperform DSI-QG significantly for MS MARCO 100K dataset in terms of MRR@3 (0.4472 vs. 0.2668). 
These results further validate that ED and RCs help the model to encode all the information about the corpus into the model parameter and SE-DSI works like a human with a knowledgeable brain.

\vspace{-3mm}
\section{Online Experiments}

Beyond the offline experiments, we conduct an online evaluation on a popular Chinese search engine, i.e., Baidu search engine. 

\vspace{-3mm}
\subsection{Task definition}

In practice, the user may specify his/her information needs through a query for official sites. 
Official sites are defined as Web pages that have been operated by universities, departments, or other administrative units. 
It does not apply to websites operated by individuals, such as students or faculty. 
For example, given a query ``北京协和医院 (Peking Union Medical College HOSP)'', the user tends to find its official site, corresponding to the site URL ``www.pumch.cn''. 
Such an authority-sensitive retrieval scenario requires high reliability and authority. 
Therefore, Baidu search sets up the site retrieval task, which is used to understand query intents on official sites, and further guide the search engine to recall relevant official sites.
Since the total number of the official site URL set is moderate, and the update frequency is lower than other retrieval scenarios, it is suitable to apply the DSI paradigm for official site retrieval. 

\vspace{-2mm}
\subsection{Datasets and evaluation metrics}
\quad \textbf{Datasets.} The official site attributes are as follows.  \begin{enumerate*}[label=(\roman*)]
 \item \textbf{Site URL} is an address for a site.
    \item \textbf{Site name} is a descriptive name that will appear in the Internet Information Services management interface.
    \item \textbf{Site Domain} is the identity of one or more site addresses. 
    \item \textbf{ICP record} is a registration name used for the Chinese Ministry of Industry and Information Technology (MIIT).
    \item \textbf{Web page} is a hypertext document on the World Wide Web. 
\end{enumerate*}
For example, for the site URL ``www.pumch.cn'', its site name is ``
北京协和医院
(Peking Union Medical College HOSP)'', the domain is ``pumch.cn'', and ICP record is ``中国医学科学院北京协和医院 (Chinese Academy of Medical Sciences and Peking Union Medical College)''. 
All data are collected from real search logs.


\textbf{Evaluation metrics.} Since the goal is to capture the positives in the top-$k$ results, we take Recall@k as evaluation metrics, where k=\{3,20\}. 
Specifically, we consider two evaluation settings for Recall, 
 \begin{enumerate*}[label=(\roman*)]
    \item \textbf{Site-level Recall@k}: the predicted site URL is completely consistent with the ground-truth site URL. 
    \item \textbf{Domain-level Recall@k}: the predicted site URL and the ground-truth site URL are in the same site domain. 
\end{enumerate*}
For example, given the ground-truth site URL ``www.pumch.cn'', if the predicted URL is ``www.pumch.cn'', it is correct on both levels. 
If the predicted site URL is ``jobs.pumch.cn'', it would be wrong at the site level, while be correct at the domain level.
We show the relative Recall ($\Delta$Recall@$k$), which is the difference value between the proposed method SE-DSI and the baseline. $\Delta$ Recall@$k > 0$ means SE-DSI is better than the baseline.

\vspace{-2mm}
\subsection{Baselines}
There are two dense retrieval methods previously used in Baidu: 
\begin{enumerate*}[label=(\roman*)]
    \item \textbf{DualEnc} is an Ernie-based\cite{sun2021ernie} dual-tower architecture model. It needs to learn a query encoder and a site encoder with (query, site attributes) pairs, where the site attributes use the site name, ICP record, and web page contents. 
    \item \textbf{SingleTow} is a single-tower method, including an Ernie-based encoder and a feed-forward layer, in which the weight is initialized with the site representations learned from DualEnc. During training, it takes the query as input, and the output logits of the feed-forward layer are passed through a softmax function, generating a probability distribution of sites. The probability of each site serves as the relevance score.
 
\end{enumerate*} During inference, DualEnc needs both queries and site attributes as input, while SingleTow only needs queries as input.




\subsection{Implementation details}
 
For model architecture, our SE-DSI is initialized with Ernie-GEN \cite{xiao2020erniegen}, an enhanced multi-flow seq2seq pre-training and fine-tuning framework.  
For the encoder of DualEnc and SingleTow, the parameters are initialized with Ernie\cite{sun2021ernie}. 
Both Ernie-GEN and Ernie are proposed by the Baidu team. 
For SingleTow, the site representation layer is randomly initialized.

For elaborative description, since some sites are not associated with web pages in practical, we directly use the unique site URLs as the docids.
For rehearsal contents, we use the leading passages and sentences of each web page for the leading-style augmentation, 
 where the number of the leading passages and sentences is 2 and 6, respectively. 
For the summarization-style augmentation method in RCs, we extract important passages and sentences from each web page, and set the number of important passages and sentences as 1 and 6, respectively. 
Specifically, we leverage the textrank4zh\cite{tzh} to implement the TextRank for the Chinese language.

To learn the associations between the site attributes and site URLs, if the site has all site attributes, we train SE-DSI$_{Doc}$ with (site name, site URL) pairs, (ICP record, site URL) pairs, (web page contents, site URL) pairs. 
Further, for SE-DSI$_{Lead}$ and SE-DSI$_{Sum}$, we replace the (web page contents, site URL) pairs with (RCs, site URL) pairs. 
To map each query to its relevant site URL, we train SE-DSI models with (query, site URL) pairs. 
All experiments are conducted on the Baidu PaddleCloud platform\cite{paddle}.
During inference, SE-DSI uses the prefix tree of sites to decode the ED with 5 beams.

\begin{table}[t]
    \caption{Online A/B experimental results under the automatic evaluation. All the values are statistically significant ($t$-test with $p < 0.05$).}
    \label{tab:online}
    \centering
    \setlength{\tabcolsep}{0.5pt}
    \renewcommand{\arraystretch}{0.9}
    \begin{tabular}{l|c|c|c|c}
        \toprule
        \multirow{2}{*}{\textbf{Methods}} &  \multicolumn{2}{c}{Site Level}& \multicolumn{2}{c}{Domain Level}\\
        & \textbf{$\Delta$ Recall@3} & \textbf{$\Delta$Recall@20} & \textbf{$\Delta$ Recall@3} & \textbf{$\Delta$ Recall@20} \\
    \midrule
    \multicolumn{5}{c}{\textit{Compared with DualEnc}} \\
    \midrule
        SE-DSI$_{Doc}$	&	+32.92\%	&+38.27\%&	+38.53\%	&+39.48\%\\
        \textbf{SE-DSI$_{Lead}$}	 &   +36.21\%&	+40.93\%&	+41.59\%	&+42.11\%\\
        \textbf{SE-DSI$_{Sum}$}	&	+36.95\%&	+42.40\%	&+42.45\%	&+42.97\%\\
        \midrule
        \multicolumn{5}{c}{\textit{Compared with SingleTow}} \\
        \midrule
        SE-DSI$_{Doc}$&	+3.41\%  & +4.60\%  & +2.32\%   & +3.45\%\\
        \textbf{SE-DSI$_{Lead}$}& +6.77\% & +7.32\%  &  +5.34\%  &	+6.13\%\\
        \textbf{SE-DSI$_{Sum}$}	& +7.41\% & +8.83\% & +6.20\%	& +6.91\% \\

        \bottomrule
    \end{tabular}
    \vspace{-1mm}

\end{table}

\vspace{-3mm}
\subsection{Online A/B experimental results}

As shown in Table \ref{tab:online}, in general, SE-DSI$_{Sum}$ outperforms DualEnc and SingleTow in terms of all metrics significantly. The reason might be that,
(1) DualEnc optimizes the model in the manner of directly matching the query and the site attributes. Therefore it needs high-quality site attributes to train the site encoder. However, many sites lack attributes, and web pages usually have noisy information, which may hurt performance. 
(2) SingleTow works better than DualEnc by a large margin. The reason may be that site attributes are encoded into the model in the form of a matrix, contributing to better interaction with the query. 
(3) For SE-DSI, the site representation is in the form of model parameters, making the query interact with global information, which is more flexible and deeper than explicit similarity functions. 
(4) SE-DSI$_{Sum}$ and SE-DSI$_{Lead}$ work better than SE-DSI$_{Doc}$, which shows that learning with important contents of the web pages facilitates the process of encoding the corpus, and further contributes to the retrieval.

\textbf{Case}. We conduct case studies to analyze the difference between SE-DSI$_{Sum}$ and baselines. 
Specifically, we take one example from the test set, and show the top-3 retrieval results. As shown in Table \ref{tab:site-retrieval-case}, we can see that: given the same query, DualEnc can not rank the ground-truth site URL in the top 3. SingleTow ranks the ground-truth at the 3-th, while our SE-DSI$_{Sum}$ ranks it at the 1st.

\begin{table}[t]
    \caption{An example of official site retrieval. Given the user query, DualEnc, SingleTow and SE-DSI$_{Sum}$ return the top-3 results. Correct results are marked bold.}
    \label{tab:site-retrieval-case}
    \setlength{\tabcolsep}{4pt}
    \renewcommand{\arraystretch}{1}
    \begin{tabular}{lcccc}
        \toprule
      \multicolumn{4}{l}{\multirow{1}{*}{\begin{minipage}{0.95\linewidth}\textbf{Query}: 北京协和医院(Peking Union Medical College HOSP)\end{minipage}}} \\ \hline
        \textbf{\#} & \textbf{DualEnc} & \textbf{SingelTow} &\textbf{SE-DSI$_{Sum}$} \\
        \hline
        1 &  hospital.pku.edu.cn &   www.bjhmoh.cn & \textbf{www.pumch.cn}\\
        2 &  www.bjmu.edu.cn &  www.pumc.edu.cn &ims.pumch.cn\\
        3 &  www.youlai.cn & \textbf{www.pumch.cn} & jobs.pumch.cn\\
      
       \bottomrule
    \end{tabular}
\end{table}

\textbf{Side-by-side comparison}. 
Besides, we also conduct a side-by-side comparison between SingleTow and the combination method of SE-DSI$_{Sum}$ and the SingleTow in terms of overall satisfaction and high-quality authority.
Human experts judge whether the combination method or the SingleTow gives better final results.
Here, the relative gain is measured with Good vs. Same vs. Bad (GSB) as
\begin{equation*}
    \Delta GSB = \frac{\#Good - \#Bad}{\#Good + \#Same + \#Bad},
\end{equation*}
where $\#Good$ (or $\# Bad$) indicates the number of queries that the combination method provides better (or worse) final results.
As shown in table \ref{tab:GSB}, we can find that 
it has achieved significant positive gains in terms of both aspects. 

\textbf{Inference speed.} We analyzed the end-to-end inference time of the retrieval phase:
\begin{enumerate*}[label=(\roman*)]
    \item Compared to DualEnc, the running speed of SE-DSI$_{Sum}$, which is proportional to the beam size, has been significantly improved by about 2.5 times. 
    \item The running speed of SE-DSI$_{Sum}$ is about the same as SingleTow, which classifies sites with one softmax operation.
    \item In general, the running speed of SE-DSI$_{Sum}$ can meet the requirements of industrial applications. 
\end{enumerate*}

\section{Related Work}

\textbf{Sparse retrieval}. The key idea of sparse retrieval methods is to utilize exact matching signals to design a relevance scoring function. Specifically, these models consider easily computed statistics (e.g., term frequency, document length, and inverse document frequency) of normalized terms matched exactly between the query and document. Among these models, BM25 \cite{bm25} is shown to be effective and is still regarded as a strong baseline of many retrieval models nowadays. 
To enhance the semantic relationships, several works utilize word embeddings as term weights ~\cite{pipeline2,deepTR}. 

\noindent \textbf{Dense retrieval}. To solve the vocabulary mismatch problem in sparse retrieval \cite{zhao2010term,furnas1987vocabulary}, many researchers turn to dense retrieval models \cite{zhan2020RepBERT,luan2021sparse}, which first learn dense representations of both queries and documents, and then approximate nearest neighbor search \cite{bentley1975multidimensional,beis1997shape} is employed to retrieve. 
Further, pre-trained models are used to enhance dense retrieval \cite{nie2020dc,khattab2020colbert}. 


\noindent \textbf{Differentiable search index}. Differentiable Search Index (DSI) \cite{DSI} is gaining increasing attention, which retrieves documents by generating their docid using a generative model.  
It presents an end-to-end solution for document retrieval tasks and allows for better exploitation of the capabilities of pre-trained  generative models.

For the docids, the original DSI proposed that the docid could be represented by a single token (atomic integers) or a string of tokens, which can be an arbitrary string or a semantic numeric string \cite{DSI}. 
Some later works followed this way to define the docids  \cite{NCI,bridging}. 
Though the semantic numeric docid enables that semantically similar documents to share prefixes, it is insufficient and implicit to reflect the semantic meaning of the document. 
This way, it is sub-optimal to map docids into a suitable semantic space. 
To further enrich the semantic information, researchers proposed to leverage Wikipedia page titles \cite{chen2022gere, chen2022corpusbrain,genre} as the docids for Wikipedia-based tasks. 
However, such methods depend on certain special document metadata. 
To mitigate this limitation, some works proposed leveraging all n-grams in a passage as its possible docid \cite{seal,chen-2023-unified}.
But it is costly to enumerate all occurrences of n-grams in the corpus. 
Here, we propose to construct EDs from documents to represent them, containing sufficient semantic information. 

\begin{table}[t]
    \caption{Human evaluation results in terms of $\Delta$GSB. All the values are statistically significant ($t$-test with $p < 0.05$).}
    \label{tab:GSB}
    \centering
    \setlength{\tabcolsep}{20pt}
    \renewcommand{\arraystretch}{1}
    \begin{tabular}{l|c}
        \toprule
        Aspect & $\Delta$GSB \\
        \midrule
        Overall satisfaction & +2.99\% \\
        High-quality and authority & +11.52\% \\
        \bottomrule
    \end{tabular}
\end{table}

For the associations between documents and docids, the original DSI model proposed to take document tokens as input and generate docids as output \cite{DSI}. 
Though simple and effective, documents of long length might be hard for the model to capture and result in poor performance.  
Later, some researchers proposed to only use multiple short pseudo queries generated from the documents as the input \cite{bridging,NCI}, and then pair them with the semantic numeric string \cite{DSI}. 
However, only encoding pseudo queries may lose some essential information.  
Differently, we propose to select multiple important parts in the document, jointly with the original document, to improve document memorization. 

\section{Conclusion}
In this work, we pointed out that designing a proper generative model to ``memorize'' the whole corpus for document retrieval remains a challenge. 
Inspired by learning strategies, we have proposed SE-DSI to advance the original DSI, which takes the input of the original document augmented with RCs containing important parts and outputs the ED with explicit semantic meanings.  
The offline experimental results on several representative retrieval datasets demonstrated the effectiveness of our SE-DSI model. 
The online evaluation again verified the value of this work.

As a novel document retrieval paradigm, the performance of DSI models remains a large room to be improved. 
In future work, we would like to focus on the following directions, (1) Scenario: the document corpus is usually dynamic in real-world search engines; (2) Architecture: there is potential in exploring to use other model architectures or yet to come larger autoregressive models;  (3) Learning: how to define learning strategies and identifiers, etc. 

\begin{acks}
This work was funded by the National Natural Science Foundation of China (NSFC) under Grants No. 62006218, the China Scholarship Council under Grants No. 202104910234, the Youth Innovation Promotion Association CAS under Grants No. 20144310, and the Lenovo-CAS Joint Lab Youth Scientist Project. 
We would like to thank the reviewers for their valuable feedback and suggestions.
\end{acks}

\newpage
\bibliographystyle{ACM-Reference-Format}
\bibliography{main}


\end{CJK*}
\end{document}